# Convergence of Local Dynamics to Balanced Outcomes in Exchange Networks

Yossi Azar*    Benjamin Birnbaum†    L. Elisa Celis‡    Nikhil R. Devanur§

Yuval Peres¶

October 30, 2018

## Abstract

Bargaining games on exchange networks have been studied by both economists and sociologists. A *Balanced Outcome* [10, 16] for such a game is an equilibrium concept that combines notions of stability and fairness. In a recent paper, Kleinberg and Tardos [14] introduced balanced outcomes to the computer science community and provided a polynomial-time algorithm to compute the set of such outcomes. Their work left open a pertinent question: *are there natural, local dynamics that converge quickly to a balanced outcome?* In this paper, we provide a partial answer to this question by showing that simple edge-balancing dynamics converge to a balanced outcome whenever one exists.

---

*azar@tau.ac.il, Microsoft Research, Redmond and Tel-Aviv University, Tel-Aviv, Israel.

†birnbaum@cs.washington.edu, University of Washington, Department of Computer Science and Engineering. Research supported by an NSF Graduate Research Fellowship

‡ecelis@cs.washington.edu, University of Washington, Department of Computer Science and Engineering. Research supported by an NPSC Graduate Research Fellowship.

§nikdev@microsoft.com, Microsoft Research, Redmond.

¶peres@microsoft.com, Microsoft Research, Redmond and University of California, Berkeley.

# 1 Introduction

Exchange networks and their properties have been well-studied by sociologists and economists, who have often made similar discoveries independently. For a sociologist, an exchange is any social interaction based on reciprocity. For an economist, an exchange is a trade that generates a surplus for the parties involved. A network of exchanges arises when there are multiple players involved in pairwise exchanges but when the potential pairings between players are limited. The object of interest in such networks is how the players bargain for the surplus generated by exchanges, and how the outcome depends on the topology of the network. In sociology, this is studied using network exchange theory [21], and in economics it is studied using cooperative game theory [18, 20].

Interaction in such exchange networks takes the form of the following bargaining game [10, 20], defined by a graph $G = (V, E)$ with edge weights $w : E \to \mathbb{Z}_+$. Each node of the graph is a player, and the weight on an edge signifies the value generated by an exchange between the two players connected by the edge. Any two players that are connected by an edge are allowed to negotiate on how to split the value on the edge. Eventually, however, each player is allowed an exchange with only one other player. An outcome of such a game corresponds to a matching $M$, which represents the pairs of players involved in an exchange, and a vector $\mathbf{x}$ describing each player's *allocation* of the value generated by the exchange. If $x_u$ is the allocation of node $u$ in the outcome, then for all edges $uv \in M$, we have $x_u + x_v = w_{u,v}$, and for all $u \notin M$, we have $x_u = 0$.

There are various notions of equilibrium that are intended to capture rational play in this game. The most basic and natural one is that of a *stable* outcome. In such an outcome, no two players have an unrealised exchange that is better for each of them than their realised exchanges; in other words, $x_u + x_v \geq w_{u,v}$ for all unmatched edges $uv$. Stability, however, is a mild restriction that does not fully capture the bargaining aspect of the game. Rochford [16], and independently, Cook and Yamagishi [10], introduced the notion of *balanced* outcomes, which has been found to match experimental data on such games quite well [8, 9]. Balanced outcomes can be seen as a generalization of Nash's bargaining solution for two players [15], so we describe that first. Suppose that two players have a dollar to split among themselves if they agree how to split it. Each player also has an alternative which is his utility in case they disagree.[1] If the alternatives of the two players are $\alpha_1$ and $\alpha_2$, then Nash's bargaining solution suggests that they split the surplus $s = 1 - (\alpha_1 + \alpha_2)$ equally if it is positive, and disagree otherwise. That is, the players get $\alpha_1 + \frac{s}{2}$ and $\alpha_2 + \frac{s}{2}$ if $s \geq 0$. For the network bargaining game, an outcome is balanced if it is stable and the value of each realised exchange is split according to Nash's bargaining solution. However, as opposed to the *exogenously* given alternatives in Nash's game, the alternatives in the network bargaining game are given *indigenously*, to be the maximum value a player can get by executing an exchange with a neighbor other than his match and offering that neighbor the same value that he is currently getting. More precisely, the alternative of a node $u$ is $\alpha_u = \max\{0, \max_{v:uv \in E \setminus M}\{w_{u,v} - x_v\}\}$. In a recent work that brought this concept into the realm of computer science, Kleinberg and Tardos [14] gave a polynomial-time algorithm to compute the set of balanced outcomes.

Kleinberg and Tardos asked a pertinent question left open by their work: "*are there natural, local dynamics that converge to a balanced outcome?*" In fact, Rochford [16] and Cook and Yamagishi [10] already define one such process. This process, which we call the *edge-balancing dynamics*, assumes that the players have determined their matches and are just negotiating on their allocations. With this matching $M$ fixed, the process works as follows: for a matched edge $uv$, let the surplus be $s_{u,v} = w_{u,v} - (\alpha_u + \alpha_v)$. If the edge is not already balanced, then rebalance it by setting $x_u$ and $x_v$ to the values suggested by Nash's bargaining solution: $x_u \leftarrow \alpha_u + \frac{s_{u,v}}{2}$ and $x_v \leftarrow \alpha_v + \frac{s_{u,v}}{2}$. Do this even if the surplus $s_{u,v}$ is negative – unless doing this would make $x_u$ or $x_v$ negative, that is, unless $\alpha_u + \frac{s_{u,v}}{2} < 0$ or $\alpha_v + \frac{s_{u,v}}{2} < 0$. In the first of these cases, set $x_u \leftarrow 0$ and $x_v \leftarrow w_{u,v}$, and in the second, set $x_u \leftarrow w_{u,v}$ and $x_v \leftarrow 0$. In this way, the allocation on each edge is set to be as close to balanced as possible while maintaining the condition that $x_u$ and $x_v$ are always non-negative and add up to $w_{u,v}$.[2]

---
[1]This is also sometimes called the disagreement point.

In order to completely specify the dynamics, one has to specify the order in which the edges are rebalanced and the initial state. For instance, if all matched edges are simultaneously rebalanced, then it is easy to show that the dynamics may cycle. On the other hand, if the dynamics start at a particular state, then it is easy to show that they converge. (More on this in the section on related results.) In order to be as general as possible, we consider an *arbitrary* order of edges and an *arbitrary* initial state. *Our main result is that the edge-balancing dynamics converge to a balanced outcome whenever one exists with matching $M$.*

In fact, it is not even clear that a fixed point of this process is a balanced outcome. This is because, at a fixed point, the surplus of an edge could be negative, or worse, it could be so negative that $\alpha_u + \frac{s_{u,v}}{2} < 0$ and we set $x_u \leftarrow 0$, and $x_v \leftarrow w_{u,v}$. In Section 3, we show examples where the edge-balancing dynamics do indeed get stuck on such an outcome. However, in each of these examples, it can be seen that there is no balanced outcome with respect to the given matching $M$. To prove our result, we need to argue that this is always the case: the dynamics can only get stuck in such a way if there is no balanced outcome with $M$. To do this, we show that if a fixed point is not a balanced outcome, then an "exploration" algorithm always finds a structure in the graph that acts as a witness to the non-existence of a balanced outcome.

The convergence proof uses a potential function argument. Given the potential function, the proof of convergence is fairly easy. However, the potential function we use is somewhat peculiar, owing to the fact that many natural potential functions do not work. This is because balancing one matched edge might lead to an increase in the imbalance of many other edges. In fact, the only special case where we know an alternate (easier) potential function is the case of a path.

In our dynamics we assume that we are given a matching $M$. One might ask why this assumption is justified. Besides seeing it as a natural starting point for research in understanding more realistic dynamics, we also make the following observation: the message passing algorithm of [4, 5, 19] for finding maximum matchings can be thought of as a first phase in which the players converge to a matching before they use our dynamics as a second phase to compute their allocation. The messages of this algorithm have a natural interpretation as offers (which is not surprising, since the algorithm has been shown to be equivalent to the auction algorithm of [6]). We elaborate on this in Appendix A.

**Other Related Work**

The bipartite version of the bargaining game was introduced by Shapley and Shubik [20], who called it the *assignment* game. Assignment games can also be thought of as a variant of the two-sided markets of Gale and Shapley [13] with transferable utilities. Assignment games have been the subject of numerous papers in game theory. The most relevant to our paper is that of Rochford [16], who also defined balanced outcomes for such games (under the name "pairwise-bargained allocations"). Rochford also considered a simplified version of our edge-balancing dynamics, and showed that if the initial state is one of two special points, then the process converges to a balanced outcome. The issue of characterizing the fixed points does not arise in Rochford's work since the initial states are such that the allocations are always guaranteed to be stable. Also, the special initial states give a natural monotonicity property that is lacking when considering arbitrary starting points.

Papers in a similar vein that consider price setting through a bargaining processes include the work by Corominas-Bosch [11]. At a meta-level, the convergence of local dynamics to a global equilibrium is a common theme. Ackerman et. al [1] showed an exponential lower bound for random best-response dynamics for the Gale-Shapley stable matching game [13]. Several papers [3, 7, 17] have studied the convergence of best-response dynamics to Nash equilibria in congestion games. In terms of structural results, Driessen [12] shows that the *kernel* is included in the *core* of an assignment game. This is in a similar spirit to one of the structural results we prove (Proposition 5).

---
[2]Actually the process as defined by Rochford and Cook and Yamigishi is simpler because they only consider initial conditions in which the surplus is always guaranteed to be non-negative.

**Future Directions**

The *rate of convergence* of the dynamics is obviously important. However, the edge-balancing dynamics may not reach an exact balanced outcome in finite time.[3] Therefore, we define a notion of sufficiently balanced outcomes, which helps in measuring the rate. This gives a bound that is exponential in the size of the graph, and linear in $1/\varepsilon$ where $\varepsilon$ is the approximation parameter. Bringing the rate down to a polynomial in the input size is a major open problem. To do this, one might also consider the process in which the edge to be balanced is picked uniformly at random.

Another direction of research is to remove the assumption that there is a fixed matching. Although this assumption is partially justified by our observation regarding the message passing algorithm of [4, 5, 19], it would be more satisfying to have dynamics that naturally mix the process of finding a matching and bargaining across the edges of the matching.

## 2 The Edge-Balancing Dynamics

Let $M$ be a matching on a graph $G = (V, E)$ with weights $w : E \to \mathbb{Z}_+$. Let $\mathbf{x} \in \mathbb{R}_+^V$ be an *allocation* on $V$. We have already defined *stable* and *balanced* outcomes. It is also useful to define stable and balanced *edges*. An edge $uv \notin M$ is *stable* if $x_u + x_v \geq w_{u,v}$ and *unstable* otherwise. An edge $uv \in M$ is balanced if its endpoints satisfy Nash's bargaining solution, that is, if the surplus is not negative ($s_{u,v} = w_{u,v} - \alpha_u - \alpha_v \geq 0$) and the values at the endpoints are $x_u = \alpha_u + s_{u,v}/2$ and $x_v = \alpha_v + s_{u,v}/2$.

We introduce the term *quasi-balanced* to denote an edge $uv \in M$ such that $x_u = \alpha_u + s_{u,v}/2$ and $x_v = \alpha_v + s_{u,v}/2$, regardless of whether or not $s_{u,v} \geq 0$. In a *quasi-balanced outcome* every matched edge is quasi-balanced. Recall that a balanced outcome is a stable outcome in which the endpoints of each edge satisfy Nash's bargaining solution. This implies in particular that in a balanced outcome, each matched edge must have positive surplus. However, it is not hard to show that the stability condition alone implies the positive surplus of the matched edges. Hence, balanced outcomes are equivalent to stable quasi-balanced outcomes; this is the characterization we will use.

We define one other type of matched edge for clarity in describing situations in which the surplus is negative. We say that an edge $uv \in M$ is *unhappy* if $\alpha_u + s_{u,v}/2 < 0$ or $\alpha_v + s_{u,v}/2 < 0$. That is, an edge is unhappy whenever the edge-balancing dynamics suggest a negative value for an endpoint. We say that vertex $v$ is *saturated* if the unhappy edge $uv \in M$ is as close to being balanced as it can get without causing $x_u$ to be negative, namely, $\alpha_u + s_{u,v}/2 < 0$, $x_v = w_{u,v}$ and $x_u = 0$. We call an edge saturated if one of its endpoints is saturated.

With these definitions, we now formally describe the edge-balancing dynamics.

---

[3]This is easily seen, even for a path on 4 nodes, since the edge-balancing dynamics may be made to always have numbers whose denominators are powers of 2, whereas a balanced outcome has values $1/3$ and $2/3$.

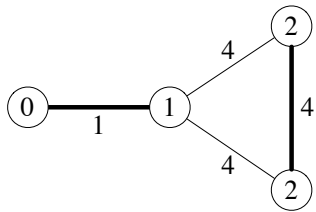
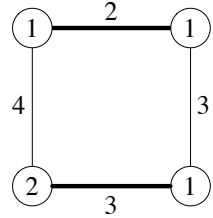

(a) Fixed point that is not quasi-balanced.

(b) Fixed point that is quasi-balanced, but not stable.

Figure 1: Examples of fixed points of the edge-balancing dynamics that are not, in fact, balanced. In Figure 1(a), the fixed point is not quasi-balanced and in Figure 1(b), the fixed point is quasi-balanced but not stable. However, it can be shown that there is no balanced outcome with either of these matchings.

---

**Edge-Balancing Dynamics**

Let $\mathbf{x}$ be an arbitrary allocation with respect to $M$.
Repeat:
    Choose any edge $uv \in M$ that is not quasi-balanced or saturated.[4]
    Let $\alpha_u$ and $\alpha_v$ be the best alternates for $u$ and $v$.
    Let $x'_u = \alpha_u + s_{u,v}/2$, and
    Let $x'_v = \alpha_v + s_{u,v}/2$.
    If $x'_u < 0$ (that is $uv$ is unhappy with $\alpha_u < \alpha_v$)
        Set $x_u \leftarrow 0$ and $x_v \leftarrow w_{u,v}$.
    Else if $x'_v < 0$ (that is $uv$ is unhappy with $\alpha_u > \alpha_v$)
        Set $x_u \leftarrow w_{u,v}$ and $x_v \leftarrow 0$.
    Else
        Set $x_u \leftarrow x'_u$ and $x_v \leftarrow x'_v$.

---

Clearly we maintain the invariant that $x_u + x_v = w_{u,v}$ for all $uv \in M$, and if $z \notin M$ then $x_z$ does not change. Thus, at any point in the dynamics, $\mathbf{x}$ is a valid allocation with respect to $M$. In a fixed point of these dynamics, every matched edge is either quasi-balanced or unhappy with a saturated endpoint, and unmatched edges may or may not be stable. Thus there can be (and are) examples of fixed points that are not quasi-balanced or stable (see Figure 1). However, our main result shows that in any such example there is no balanced outcome on matching $M$.

**Theorem 1.** *If there exists a balanced outcome on $M$, then any fixed point of the edge-balancing dynamics is balanced.*

This theorem is proved in Section 3. In Section 4, we prove that the dynamics converge to a fixed point.

**Theorem 2.** *For any initial allocation $\mathbf{x}$ and any matching $M$, the edge-balancing dynamics converge to a fixed point.*

## 3 Fixed Point Characterization

Since a stable outcome must be on a maximum weight matching [14, 20], a balanced outcome must also be on a maximum weight matching. It is also known that if $M$ is maximum, then a balanced outcome with $M$ exists if and only if a stable outcome on $G$ exists [14, 16]. Hence, a balanced outcome exists on a matching $M$ if and only if $M$ is a maximum matching and there exists a stable outcome on $G$. A straightforward duality argument [14] shows that there exists a stable outcome on $G$ if and only if a maximum fractional matching on $G$ is integral. Hence, we have the following fact.

---
[4]We assume the natural non-starvation condition that each edge is considered an infinite number of times.

**Fact 3.** *There exists a balanced outcome on matching $M \subseteq G$ if and only if $M$ is a maximum fractional matching.*

Note that in Figure 1, neither of the two matchings are maximum fractional matchings.

We will prove Theorem 1 in two steps. The first step is to show that when a balanced outcome exists, a fixed point of the edge-balancing dynamics is quasi-balanced.

**Proposition 4.** *If $M$ is a maximum fractional matching and $\mathbf{x}$ is a fixed point of the edge-balancing dynamics, then $\mathbf{x}$ has no unhappy edges (and is therefore quasi-balanced).*

The second step is to show that, given these same conditions, a quasi-balanced allocation is stable.

**Proposition 5.** *If $M$ is a maximum fractional matching and $\mathbf{x}$ is quasi-balanced, then $\mathbf{x}$ contains no unstable edges (and is therefore balanced).*

Combined with Fact 3, these two propositions prove Theorem 1.

Proposition 5 is of independent interest because it shows that although quasi-balanced is a weaker notion than balanced, the two are equivalent on matchings that allow balanced outcomes. In fact, we can prove a generalization that is motivated by practical concerns. Recall that although the edge-balancing dynamics converge, they may not reach a fixed point in finite time. However, once parties are sufficiently satisfied they may not want to negotiate further. Thus we consider a notion of sufficiently quasi-balanced or sufficiently stable.

We say that allocation $\mathbf{x}$ is $\varepsilon$-quasi-balanced if the surplus is split evenly, to within an additive constant of $\varepsilon$; namely every edge $uv \in M$ satisfies $|(x_u - \alpha_u) - (x_v - \alpha_v)| \leq \varepsilon$. Similarly, an allocation is $\delta$-stable if no two unmatched players have an unrealized exchange that is more than $\delta$ better for them; namely $x_u + x_v \geq w_{u,v} - \delta$ for each edge $uv \notin M$. In light of Proposition 5, one might ask if there is a similar relationship between $\varepsilon$-quasi-balanced and $\delta$-stable. We settle this question in the affirmative in the following proposition.

**Proposition 6.** *If $M$ is a maximum fractional matching and $\mathbf{x}$ is an $\varepsilon$-quasi-balanced allocation on $M$, then $\mathbf{x}$ is $(n\varepsilon)$-stable, where $n = |V(G)|$.*

The proof of Proposition 6 follows similar ideas as the proof of Proposition 5. We present this proof in Appendix B, and show this bound is tight to within a constant factor.

The proofs of Propositions 4 and 5 share a common technique for proving the contrapositive. We assume there exists an edge $uv$ that is unhappy (in the case of Proposition 4) or an edge $uv$ that is unstable (in the case of Proposition 5). Starting from this edge, we explore the graph along matched edges and best alternatives and show that this exploration must terminate by finding a structure with properties that imply $M$ is not a maximum fractional matching. This exploration algorithm is shown below. To ease notation, we typically write $w_{u_i,u_j}$ and $x_{u_i}$ simply as $w_{i,j}$ and $x_i$.

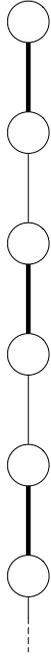
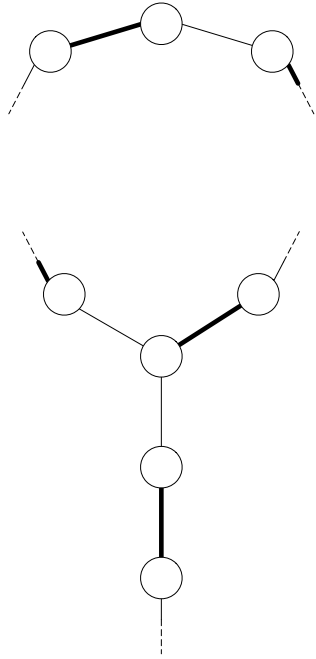
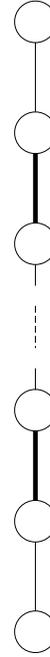

(a) *Capped*: at least one endpoint is matched. The other endpoint can take any form.

(b) *Lollypop*: at least one endpoint forms an alternating (even length) cycle. The other endpoint can take any form.

(c) *Augmenting path*: both endpoints are unmatched.

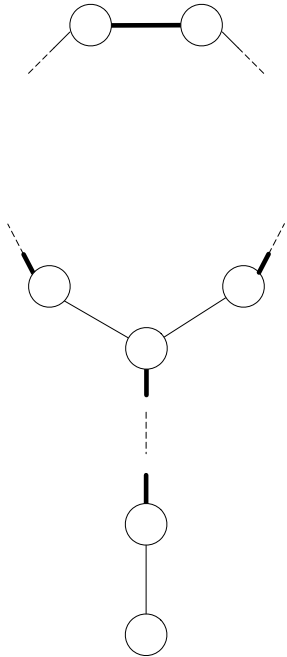
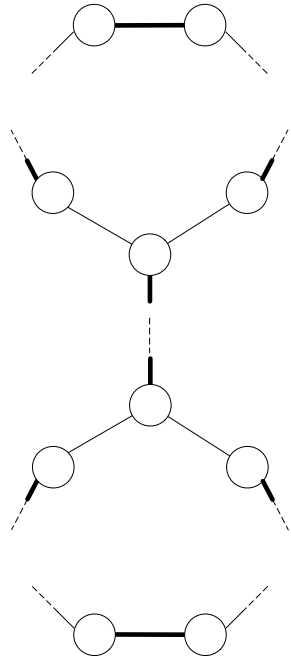
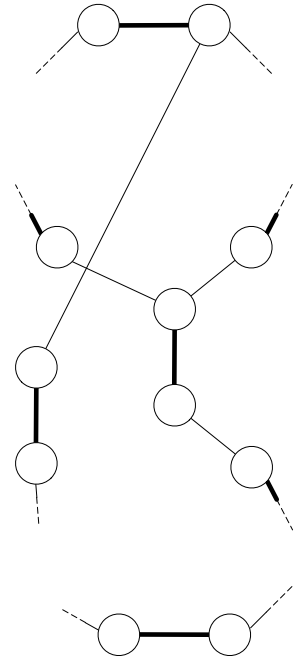

(d) *Flower*: one endpoint forms a blossom (odd cycle), and the other endpoint is unmatched.

(e) *Bicycle*: both endpoints form disjoint blossoms.

(f) *Pretzel*: both endpoints form blossoms which are not disjoint.

Figure 2: The exploration algorithm finds a subgraph $H$ that can be classified in one of these six ways. Matched edges are depicted in bold.

```
Exploration Algorithm
Choose u_0 ∈ V.
Let S = ∅ be the set of explored vertices.
For i = 0, 1, 2, 3, . . .
    If u_i ∈ S, break.
    Else,
        S ← S ∪ {u_i}.
        If i is even:
            If u_i ∉ M, break.
            Else, let u_{i+1} = M(u_i).
        Else (i is odd):
            If α_i = 0, break.
            Else, choose u_{i+1} ∼ u_i such that α_i = w_{i,i+1} − x_{i+1}.
For i = 0, −1, −2, −3, . . .
    If i < 0 and u_i ∈ S, break.
    Else,
        S ← S ∪ {u_i}.
        If i is odd:
            If u_i ∉ M, break.
            Else, let u_{i−1} = M(u_i).
        Else (i is even):
            If α_i = 0, break.
            Else, choose u_{i−1} ∼ u_i such that α_i = w_{i,i−1} − x_{i−1}.
```

Note that by the definition of best alternate $\alpha_i$, when $\alpha_i > 0$ a vertex $u_{i'} \sim u_i$ exists such that $\alpha_i = w_{i,i'} - x_{i'}$ as desired. Thus, this dynamics are well defined. Let $H$ be the subgraph formed by the set of vertices $S$ and the edges travelled in their discovery. Because both directions of the exploration terminate if a vertex in $S$ is rediscovered, the subgraph $H$ can be classified in one of the six ways shown in Figure 2.

Because the proof of Proposition 5 is simpler than the proof of Proposition 4 and more clearly illustrates our exploration technique, we present it first.

## Quasi-Balanced Outcomes are Stable

The proof of Proposition 5 makes use of the following lemma, which is proved in Appendix C. Note that the structures in this lemma played a prominent role in [14] as well.

**Lemma 7.** *Let $G$ be a graph and let $(M, \mathbf{x})$ be an outcome on $G$. If there is a subgraph $H \subseteq G$ that is an augmenting path, alternating cycle, flower, or bicycle (see Figure 3) such that*

- *for each $uv \in H \cap M$, $x_u + x_v = w_{uv}$, and*
- *for each $uv \in H \setminus M$, $x_u + x_v < w_{uv}$,*

*then there exists a (possibly fractional) matching on $G$ with weight strictly greater than $M$.*

We now proceed with the proof of the proposition.

*Proof of Propositon 5.* We prove the contrapositive. Assume there exists an unstable edge $uv \in M$. If neither $u$ nor $v$ are matched, then $M$ is not maximum. Without loss of generality, assume $u$ is matched and let $u_0 = u$. Let $u_\ell, \ldots, u_0, \ldots, u_r$ be the set of vertices discovered by the exploration algorithm, and let $H \subseteq G$ be the subgraph of vertices and edges traversed. We first prove that every edge $uv \in H \setminus M$ is unstable.

Specifically, we prove by induction that for $0 \leq i \leq \lfloor r/2 \rfloor$, we have $x_{2i-1} + x_{2i} < w_{2i-1,2i}$, and an equivalent argument holds for $0 \geq i \geq \lceil (\ell+1)/2 \rceil$. To prove the base case we show $u_{-1}u_0$ is unstable. Note $\alpha_0 \geq w_{v0} - x_v$, which implies $x_0 - \alpha_0 \leq x_0 + x_v - w_{v0} < 0$. Specifically, $0 < \alpha_0$, so $u_{-1}$ exists and $\alpha_0 = w_{-1,0} - x_{-1}$. Thus, $x_0 - (w_{-1,0} - x_{-1}) = x_0 - \alpha_0 < 0$, or equivalently $x_0 + x_{-1} < w_{-1,0}$. Hence $u_{-1}u_0$ is unstable.

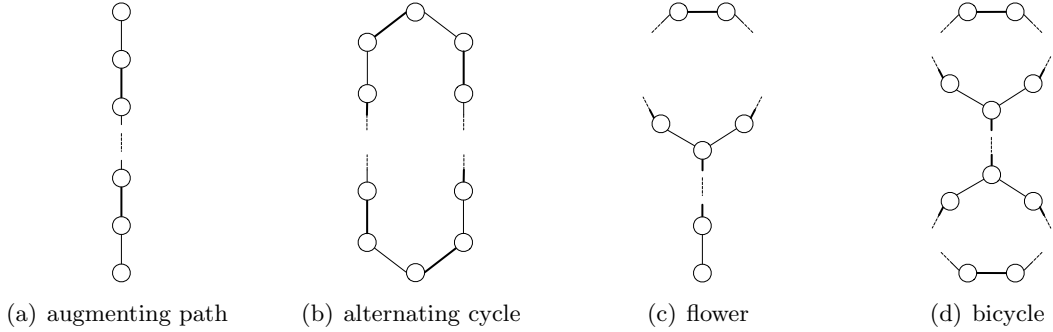

(a) augmenting path  (b) alternating cycle  (c) flower  (d) bicycle

Figure 3: Witnesses that prove that $M$ is not a maximum fractional matching.

Now let us assume $u_{2i-3}u_{2i-2}$ is an unmatched edge that is unstable and let us show that $u_{2i-1}u_{2i}$ is unstable. Since $u_{2i-2}u_{2i-1}$ is a matched edge, it is by assumption quasi-balanced. Also, since $u_{2i-3} \sim u_{2i-2}$, it follows that $\alpha_{2i-2} \geq w_{2i-3,2i-2} - x_{2i-3}$. Hence, $x_{2i-1} + x_{2i} = x_{2i-1} + w_{2i-1,2i} - \alpha_{2i-1} = x_{2i-2} - \alpha_{2i-2} + w_{2i-1,2i} \leq x_{2i-2} - w_{2i-3,2i-2} + x_{2i-3} + w_{2i-1,2i} < w_{2i-1,2i}$, and thus $u_{2i-1}u_{2i}$ is unstable. We conclude that every edge $uv \in H \setminus M$ is unstable.

Now consider $H$, and recall it takes the form of one of the six structures in Figure 2. We consider each case, and show our desired conclusion. Since every unmatched edge in $H$ is unstable, we know $\alpha_i > x_i \geq 0$ for all $i$. Thus, $H$ cannot be capped (Figure 2(a)). Now suppose that $H$ is a lollypop (Figure 2(b)) or a pretzel (Figure 2(f)). If $H$ is a lollypop then by definition it contains an alternating cycle. Alternately, if $H$ is a pretzel then a simple parity argument shows that one of its cycles must have even length, and must therefore be alternating.

Hence, $H$ is either an augmenting path (Figures 2(c)), a flower (Figures 2(d)), a bicycle (Figure 2(e)), or contains an alternating cycle. Since each unmatched edge is unstable, we apply Lemma 7 to conclude that $M$ is not a maximum fractional matching, thus proving the contrapositive. □

### Fixed Points are Quasi-Balanced

To prove that fixed points are quasi-balanced, we begin with the following lemma.

**Lemma 8.** *Let $\mathbf{x}$ be a fixed point of the edge-balancing dynamics on a maximum matching $M$, and let $u_0 u_1 \in M$ be an unhappy edge with $u_1$ saturated. Consider the vertices $u_0, u_1, \ldots, u_r$ discovered by the first part of the exploration algorithm. If $\ell \leq r$, then for any odd $\ell$:*

1. *For any $j$ such that $0 \leq j < \lfloor \ell/2 \rfloor$, we have*

$$x_\ell + \sum_{i=j+1}^{\lfloor \ell/2 \rfloor} w_{2i-1,2i} > x_{2j+1} + \sum_{i=j+1}^{\lfloor \ell/2 \rfloor} w_{2i,2i+1} .$$

2. $x_\ell > 0$.
3. $x_\ell - \alpha_\ell \leq x_{\ell-1} - \alpha_{\ell-1}$.
4. $\alpha_\ell > x_\ell$.
5. *For any odd $\ell' < \ell$, we have $\alpha_\ell - x_\ell \geq \alpha_{\ell'} - x_{\ell'}$.*

*Alternatively, for any even $\ell$:*

6. *For $\ell \geq 2$, $w_{\ell-1,\ell} > x_{\ell-1} + x_\ell$.*
7. *For $\ell \geq 2$, we have $\alpha_\ell > x_\ell$.*

8. For any $j$ such that $0 \leq j < \ell/2$, we have

$$x_\ell + \sum_{i=j}^{\ell/2-1} w_{2i,2i+1} < \sum_{i=j+1}^{\ell/2} w_{2i-1,2i} \ .$$

*Proof.* We prove this lemma by joint induction on $\ell$. When $\ell = 0$, Lemma 8(6) and Lemma 8(7) do not apply (note that their proofs when $\ell = 2$ only rely on $\ell = 1$), and Lemma 8(8) holds vacuously. Additionally, when $\ell = 1$, Lemma 8(1) holds vacuously; Lemma 8(2) holds because $x_1 = w_{0,1} > 0$; Lemma 8(3) holds because $x_0 = 0$ and $\alpha_1 - \alpha_0 > w_{0,1} = x_1$; Lemma 8(4) holds because $\alpha_1 > x_1 + \alpha_0$ and $\alpha_0 \geq 0$; and finally, Lemma 8(5) holds vacuously. Let us assume the lemma holds for for all $0, \ldots, \ell - 1$ where $\ell - 1 < r$. We now show that the claims hold for $\ell$, proving the inductive step.

First consider the case where $\ell$ is odd, and denote $\ell = 2k + 1$:

**Lemma 8(1):** Let $0 \leq j < k$. We have

$$\begin{aligned}
x_\ell + \sum_{i=j+1}^{k} w_{2i-1,2i} &= x_\ell + w_{\ell-2,\ell-1} + \sum_{i=j+1}^{k-1} w_{2i-1,2i} \\
&= (w_{\ell-1,\ell} - x_{\ell-1}) + w_{\ell-2,\ell-1} + \sum_{i=j+1}^{k-1} w_{2i-1,2i} \quad (u_{\ell-1}u_\ell \in M) \\
&> w_{\ell-1,\ell} + x_{\ell-2} + \sum_{i=j+1}^{k-1} w_{2i-1,2i} \quad \text{(Lemma 8(6), IH)} \\
&\geq w_{\ell-1,\ell} + x_{2j+1} + \sum_{i=j+1}^{k-1} w_{2i,2i+1} \quad \text{(Lemma 8(1), IH)} \\
&= x_{2j+1} + \sum_{i=j+1}^{k} w_{2i,2i+1} \ .
\end{aligned}$$

Note that while we do use Lemma 8(6) for $\ell - 1$, we know that $\ell \geq 3$, so this does not present a problem.

**Lemma 8(2):** We have

$$\begin{aligned}
x_\ell &> x_1 + \sum_{i=1}^{k} w_{2i,2i+1} - \sum_{i=1}^{k} w_{2i-1,2i} \quad \text{(Lemma 8(1), with } j = 0\text{)} \\
&= \sum_{i=0}^{k} w_{2i,2i+1} - \sum_{i=1}^{k} w_{2i-1,2i} \quad (x_1 = w_{0,1}) \\
&\geq 0 \ . \quad \text{(see below)}
\end{aligned}$$

Note that the edges $u_0 u_1$ and $u_{\ell-1} u_\ell$ are both matched. Thus the last inequality follows since $M$ is a maximum matching.[5]

**Lemma 8(3):** Because **x** is a fixed point, $u_{\ell-1}u_\ell$ is either balanced or unhappy with an endpoint saturated. By Lemma 8(2), it cannot be unhappy with $u_{\ell-1}$ saturated. Thus, it is either balanced and $x_\ell - \alpha_\ell = x_{\ell-1} - \alpha_{\ell-1}$ or it is unhappy with $u_\ell$ saturated. In the latter case, note that $x_\ell + x_{\ell-1} = w_{\ell,\ell-1} < a_\ell - a_{\ell-1}$ by definition. Since $x_{\ell-1} = 0$, we have $x_\ell - \alpha_\ell < x_{\ell-1} - \alpha_{\ell-1}$, as desired.

---

[5] The trivial proof of this statement is included in Lemma 10 in Appendix C for completeness.

**Lemma 8(4):** From Lemma 8(3), we have $\alpha_\ell \geq x_\ell - x_{\ell-1} + \alpha_{\ell-1}$. By Lemma 8(7) for $\ell - 1$ we know $\alpha_{\ell-1} > x_{\ell-1}$. Hence $\alpha_\ell > x_\ell$.

**Lemma 8(5):** First consider $\ell' = \ell - 2$. By Lemma 8(3), we have $\alpha_\ell - x_\ell \geq \alpha_{\ell-1} - x_{\ell-1}$. Since $u_{\ell-2} \sim u_{\ell-1}$, we have $\alpha_{\ell-1} \geq w_{\ell-1,\ell-2} - x_{\ell-2}$. Hence $\alpha_\ell - x_\ell \geq w_{\ell-1,\ell-2} - x_{\ell-2} - x_{\ell-1} = \alpha_{\ell-2} - x_{\ell-2}$ by our choice of $u_{\ell-1}$. Thus, $\alpha_\ell - x_\ell \geq \alpha_{\ell-2} - x_{\ell-2}$. By the induction hypothesis, this holds for all odd $\ell' < \ell$ as desired.

Now consider the case where $\ell$ is even, and denote $\ell = 2k$:

**Lemma 8(6):** By our choice of $u_\ell$, we know $w_{\ell-1,\ell} = \alpha_{\ell-1} + x_\ell$. By Lemma 8(4), we have $\alpha_{\ell-1} > x_{\ell-1}$. Thus $w_{\ell-1,\ell} > x_{\ell-1} + x_\ell$ as desired.

**Lemma 8(7):** Because $u_{\ell-1} \sim u_\ell$ and $u_{\ell-1}u_\ell \notin M$, we have $\alpha_\ell \geq w_{\ell-1,\ell} - x_{\ell-1}$. By Lemma 8(6) we have $\alpha_\ell > x_\ell$ as desired.

**Lemma 8(8):** We have

$$x_\ell + \sum_{i=j}^{k-1} w_{2i,2i+1} = x_\ell + x_{\ell-1} + x_{\ell-2} + \sum_{i=j}^{k-2} w_{2i,2i+1}$$

$$\leq \sum_{i=j+1}^{k-1} w_{2i-1,2i} + x_{\ell-1} + x_\ell \qquad \text{(see below)}$$

$$< w_{\ell-1,\ell} + \sum_{i=j+1}^{k-1} w_{2i-1,2i} \qquad \text{(Lemma 8(6))}$$

$$= \sum_{i=j+1}^{k} w_{2i-1,2i} \ .$$

The first inequality follows by the inductive hypothesis of Lemma 8(8) if $\ell = 2k \geq 4$. If $\ell = 2k = 2$, it follows because $x_{\ell-2} = x_0 = 0$.

Note that each case of the above lemma relies only on previous cases, or on the inductive hypothesis. Thus this concludes the proof. □

We can now proceed with the proof of the proposition.

*Proof of Proposition 4.* We prove the contrapositive. Recall that the only unhappy edges that appear in a fixed point must be saturated. Suppose $u_0u_1 \in M$ is unhappy with $u_1$ saturated. Consider the structure formed by the vertices $u_0, u_1, \ldots, u_r$ of the exploration algorithm started at $u_0$. Note that if the algorithm ends because $u_r$ is a previously labeled vertex, then for some $s \geq 0$ the sequence $u_s, \ldots, u_r$ forms an even alternating cycle or a blossom (odd cycle). Alternately, the algorithm ends with $u_r$ either capped or unmatched. We examine each possible case:

- If $u_r$ is capped (Figure 2(a)): From Lemma 8(4) we know that $\alpha_i > x_i \geq 0$ for all odd $i \geq 0$. Hence, by the definition of the exploration algorithm this cannot occur.

- If $u_r$ is unmatched: By Lemma 8(8) with $j = 0$ and $\ell = r$ we know

$$\sum_{i=0}^{r/2-1} w_{2i,2i+1} < \sum_{i=1}^{r/2} w_{2i-1,2i} \ .$$

Since $u_0$ is matched to $u_1$, and $u_r$ is not matched at all, $M$ is not a maximum matching.[5]

- If $u_s, \ldots, u_r$ forms an even alternating cycle for some $0 \leq s < r$ (Figure 2(b)):

First suppose $u_s = u_0$, so $u_0, \ldots, u_r$ forms a single alternating cycle, and $r$ is even. By Lemma 8(5), $\alpha_{r-1} - x_{r-1} \geq \alpha_1 - x_1$. Since $x_0 = 0$, and $u_r = u_0$, we know $\alpha_{r-1} = w_{r-1,0}$. Because $u_{r-1}$ is adjacent to $u_0$, we have $\alpha_0 \geq w_{r-1,0} - x_{r-1}$. Hence $\alpha_0 \geq \alpha_1 - x_1$, and $x_1 = w_{0,1}$ implies $\alpha_1 - \alpha_0 \leq w_{0,1}$, contradicting the fact that $u_0 u_1$ is unhappy with $u_1$ saturated.

Otherwise, we have an alternating cycle where $0 < s$. Note that $r$ is even since any preceding $u_s$ is already matched. If $r$ is even, then $s$ is also even, and by Lemma 8(8) with $j = s/2$ and $\ell = r$, we know

$$\sum_{i=s/2}^{r/2-1} w_{2i,2i+1} < \sum_{i=s/2+1}^{r/2} w_{2i-1,2i} .$$

Thus, $M$ is not a maximum matching.[5]

- If $u_s, \ldots, u_r$ form a blossom: We know that $r = 2k$ must be even and $s = 2k' - 1$ (where $k' < k$) must be odd. Note that although $u_s \ldots u_r$ forms a blossom and $u_0 u_1 \ldots u_s$ forms a stem, the bottommost edge is matched. Hence this is not equivalent to a flower, and Lemma 7 does not apply. Nevertheless, we can argue in a similar way to its proof by constructing a fractional **y** matching of weight higher than $M$. This is done as follows.

$$y_e = \begin{cases} 1 & \text{if } e = u_i u_{i+1} \notin M \text{ for } 1 \leq i \leq 2k' - 2 \\ \frac{1}{2} & \text{if } e = u_i u_{i+1} \text{ for } 2k' - 1 \leq i \leq 2k - 1 \\ 1 & \text{if } e \in M \text{ but } e \text{ not in explored structure} \\ 0 & \text{otherwise} \end{cases}.$$

This places weight $1/2$ on the edges in the blossom, weight 1 the edges of the stem which are not in $M$. We wish to show that

$$\sum_{i=1}^{k'-1} w_{2i-1,2i} + \frac{1}{2} \sum_{i=s}^{r-1} w_{i,i+1} > \sum_{i=0}^{k-1} w_{2i,2i+1} ,$$

or equivalently

$$2 \sum_{i=1}^{k'-1} w_{2i-1,2i} + \sum_{i=k'}^{k} w_{2i-1,2i} > 2 \sum_{i=0}^{k'-1} w_{2i,2i+1} + \sum_{i=k'}^{k-1} w_{2i,2i+1} .$$

We have

$$2 \sum_{i=1}^{k'-1} w_{2i-1,2i} + \sum_{i=k'}^{k} w_{2i-1,2i}$$

$$> 2 \sum_{i=1}^{k'-1} w_{2i-1,2i} + \left( \sum_{i=k'}^{k-1} w_{2i-1,2i} + x_{2k-1} \right) + x_{2k} \quad \text{(Lemma 8(6))}$$

$$> 2 \sum_{i=1}^{k'-1} w_{2i-1,2i} + \left( \sum_{i=k'}^{k-1} w_{2i,2i+1} + x_{2k'-1} \right) + x_{2k} \quad \text{(Lemma 8(1), } \ell = 2k-1)$$

$$= 2 \left( \sum_{i=1}^{k'-1} w_{2i-1,2i} + x_{2k'-1} \right) + \sum_{i=k'}^{k-1} w_{2i,2i+1} \quad (u_{2k} = u_{2k'-1})$$

$$= 2 \left( \sum_{i=1}^{k'-1} w_{2i-1,2i} - x_{2k'-2} + w_{2k'-2,2k'-1} \right) + \sum_{i=k'}^{k-1} w_{2i,2i+1} \quad (u_{2k'-2,2k'-1} \text{ matched})$$

$$\geq 2 \left( \sum_{i=0}^{k'-2} w_{2i,2i+1} + w_{2k'-2,2k'-1} \right) + \sum_{i=k'}^{k-1} w_{2i,2i+1} \quad \text{(see below)}$$

$$= 2 \sum_{i=0}^{k'-1} w_{2i,2i+1} + \sum_{i=k'}^{k-1} w_{2i,2i+1} .$$

The last inequality follows from Lemma 8(8) if $k' \geq 2$, and if $k' = 1$, it follows because $x_{2k'-2} = x_0 = 0$. Hence, we conclude that $M$ is not a maximum weighted fractional matching.

Thus, in all cases $M$ is not a maximum fractional matching, and we have proved the contrapositive. □

## 4 Convergence

In this section, we prove Theorem 2. Note the statement of the theorem holds for any $G = (V, E)$ and $M$, although as previously stated, if $M$ is not a maximum fractional matching then the point we converge to is not balanced. In this section, let $n = |V|$ and $m = |E|$.

*Proof of Theorem 2.* Let $W = \max\{w_{uv} : uv \in E\}$, so a valid allocation $\mathbf{x}$ is a point in $[0, W]^n$. Consider the multiset $S = \{x_v : v \in V\} \cup \{x_u + x_v - w_{uv} : uv \in E\}$. We call the values in $S$ the *slacks* of $\mathbf{x}$. Let $\mathbf{s} = \mathbf{s}(\mathbf{x}) \in [-W, 2W]^{n+m}$ be the vector obtained by sorting the values in $S$ in non-decreasing order. Define the potential function $\Phi : [-W, 2W]^{n+m} \to \mathbb{R}$ by

$$\Phi(\mathbf{s}) = \sum_{i=1}^{n+m} 2^{-i} s_i.$$

We first show the value of $\Phi$ converges as we run the edge-balancing dynamics, and then prove this implies the convergence of $\mathbf{x}$. We know the entries of $\mathbf{s}$ are at most $2W$, so $\Phi(\mathbf{s}) \leq 2W$ for any such slack vector. We now show that $\Phi$ is monotonically increasing as we bargain, and thus converges.

Assume $\mathbf{x}$ is not fixed, and let $\mathbf{x}'$ be the allocation obtained from one step of the balancing dynamics. Let $uv \in M$ be the edge selected for this step, and suppose that $x'_u = x_u + \varepsilon$ (and thus $x'_v = x_v - \varepsilon$) for some $\varepsilon > 0$. Note that the slacks that change are those corresponding to edges adjacent to $u$ or $v$ (except for $uv$), and the vertices $u$ and $v$. Additionally, each such slack changes by exactly $\varepsilon$, and thus $\|\mathbf{x} - \mathbf{x}'\|_\infty = \|\mathbf{s} - \mathbf{s}'\|_\infty = \varepsilon$.

For any $au \in E$ where $a \neq v$ we have $x_u - \alpha_u \leq x_u - (w_{a,u} - x_a) = x_a + x_u - w_{a,u}$ and $x_u - \alpha_u \leq x_u$. Similarly, for any $bv \in E$ where $b \neq u$, we have $x_v - \alpha_v \leq x_v - (w_{b,v} - x_b) = x_b + x_v - w_{b,v}$ and $x_v - \alpha_v \leq x_v$. Note that since $x_u < x'_u$ we know that $x'_u \neq 0$. Thus, either $|\alpha_u - \alpha_v| \leq w_{u,v}$, or $\alpha_u - \alpha_v > w_{u,v}$. In either case, $x_u < x'_u \leq \frac{1}{2}(w_{u,v} + \alpha_u - \alpha_v)$. Since $x_u = w_{u,v} - x_v$, we have $x_u + (w_{u,v} - x_v) < w_{u,v} + \alpha_u - \alpha_v$, or equivalently $x_u - \alpha_u < x_v - \alpha_v$. Hence, before the balancing step, the minimum slack that will change is $x_u - \alpha_u$. (Note that when $u$ has degree 1 this is simply $x_u$.) Similarly, after the balancing step, we can show that $x'_u - \alpha_u$ is the minimum slack that changed. Hence, $\Phi(\mathbf{s}') - \Phi(\mathbf{s}) \geq (x'_u - x_u) 2^{-(m+n)} = \varepsilon 2^{-(m+n)}$. We conclude that $\Phi$ is increasing, and thus convergent.

However, $\Phi$ is simply a function of the vector $\mathbf{x}$. Its convergence does not necessarily imply the convergence of $\mathbf{x}$. This is not hard to show, however. First, since $\Phi$ converges, the quantity $2^{n+m}\Phi$ must also converge and is therefore Cauchy. Because $2^{n+m}\Phi(\mathbf{s}') - 2^{n+m}\Phi(\mathbf{s}) \geq \varepsilon = \|\mathbf{x}' - \mathbf{x}\|_\infty$, the vector $\mathbf{x}$ is Cauchy under the $\ell_\infty$ norm. Therefore, as the edge-balancing dynamics proceed, $\mathbf{x}$ converges to a point in $[0, W]^n$.

Thus the edge-balancing dynamics converge regardless of the sequence of edges chosen. However, without a non-starvation condition, it is possible that we will not converge to a fixed point of the dynamics. Despite this fact, it suffices for every edge to be considered an infinite number of times. Assume for the sake of contradiction that we converge to a point $\mathbf{z}$ that is not fixed. Since $\Phi$ is increasing, we know $\Phi(\mathbf{x}) \leq \Phi(\mathbf{z})$ for all intermediate $\mathbf{x}$. Additionally, since $\mathbf{z}$ is not a fixed point, there exists some edge $uv$ that is not quasi-balanced or saturated. Let $\varepsilon$ be such that $uv$ is $\varepsilon$-quasi-balanced in allocation $\mathbf{z}$. Since both $\Phi$ and $\mathbf{x}$ converge, we can proceed until $uv$ is at least $\varepsilon/2$ unbalanced and $\Phi(\mathbf{z}) - \Phi(\mathbf{x}) < \varepsilon 2^{-(n+m)}/4$ for all remaining allocations $\mathbf{x}$. Since the non-starvation condition ensures that $uv$ be balanced eventually, at some point $\Phi(\mathbf{x})$ must increase by at least $\varepsilon 2^{-(n+m)}/2$, and $\Phi(\mathbf{x}) > \Phi(\mathbf{z})$, which gives a contradiction. Hence we converge to a fixed point of the dynamics. □

Recall that while the dynamics always converge, they may not reach a fixed point in a finite amount of time. However, if we fix some $\varepsilon$ and modify the edge-balancing dynamics so that edges only balance if they are not $\varepsilon$-quasi-balanced or $\varepsilon$-close to saturated, then $\Phi$ must increase by at least $\varepsilon 2^{-(n+m)}$ in each step. In addition to preventing starvation, this shows a convergence time of $2W\varepsilon \cdot 2^{n+m}$ for these dynamics. We do not know if this bound is tight, and present this as a significant open problem.

# References

[1] Heiner Ackermann, Paul W. Goldberg, Vahab S. Mirrokni, Heiko Röglin, and Berthold Vöcking. Uncoordinated two-sided matching markets. In *ACM Conference on Electronic Commerce*, pages 256–263, 2008.

[2] Bengt Aspvall and Yossi Shiloach. A polynomial time algorithm for solving systems of linear inequalities with two variables per inequality. *SIAM Journal on Computing*, 9(4):827–845, 1980.

[3] Baruch Awerbuch, Yossi Azar, Amir Epstein, Vahab S. Mirrokni, and Alexander Skopalik. Fast convergence to nearly optimal solutions in potential games. In *ACM Conference on Electronic Commerce*, pages 264–273, 2008.

[4] M. Bayati, D. Shah, and M. Sharma. Maximum weight matching via max-product belief propagation. In *International Symposium on Information Theory*, pages 1763–1767, 2005.

[5] Mohsen Bayati, Christian Borgs, Jennifer Chayes, and Riccardo Zecchina. On the exactness of the cavity method for weighted $b$-matchings on arbitrary graphs and its relation to linear programs. *Journal of Statistical Mechanics: Theory and Experiment*, 2008.

[6] Dimitri P. Bertsekas. Auction algorithms. In *Encyclopedia of Optimization*, pages 128–132. 2009.

[7] Steve Chien and Alistair Sinclair. Convergence to approximate nash equilibria in congestion games. In *SODA*, pages 169–178, 2007.

[8] K. S. Cook and R. M. Emerson. Power, equity, and commitment in exchange networks. *American Sociological Review*, 43(5):721–739, Oct 1978.

[9] K. S. Cook, R. M. Emerson, M. R. Gillmore, and T. Yamagishi. The distribution of power in exchange networks: Theory and experimental results. *American Journal of Sociology*, 89:275–305, 1983.

[10] K. S. Cook and T. Yamagishi. Power in exchange networks: A power-dependence formulation. *Social Networks*, 14:245–265, 1992.

[11] Margarida Corominas-Bosch. Bargaining in a network of buyers and sellers. *Journal of Economic Theory*, 115(1):35 – 77, 2004.

[12] Theo S.H. Driessen. A note on the inclusion of the kernel in the core of the bilateral assignment game. *International Journal of Game Theory*, 27, 1998.

[13] D. Gale and L. S. Shapley. College admissions and the stability of marriage. *The American Mathematical Monthly*, 69(1):9–15, 1962.

[14] Jon Kleinberg and Eva Tardos. Balanced outcomes in social exchange networks. *Symposium on Theory of Computing*, 2008.

[15] John Nash. The bargaining problem. *Econometrica*, 18:155–162, 1950.

[16] Sharon C. Rochford. Symmetrically pairwise-bargained allocations in an assignment market. *Journal of Economic Theory*, 34(2):262–281, 1984.

[17] R.W. Rosenthal. A class of games possessing pure-strategy Nash equilibria. *International Journal of Game Theory*, 2:65–67, 1973.

[18] Alvin E. Roth and Marilda A. Oliveira. *Two-sided matching: A study in game-theoretic modeling and analysis.* 1990.

[19] Sujay Sanghavi, Dmitry M. Malioutov, and Alan S. Willsky. Linear programming analysis of loopy belief propagation for weighted matching. *NIPS*, 2007.

[20] Lloyd S. Shapley and Martin Shubik. The assignment game I: The core. *International Journal of Game Theory*, 1(2):111–130, 1972.

[21] D. Willer, editor. *Network Exchange Theory*. Praeger, 1999.

## A  The Matching BP Algorithm as a Bargaining Process

In this section we describe a distributed algorithm for computing a maximum matching on a graph, and show that it can be viewed naturally as a bargaining process in the network exchange setting. Hence, this algorithm can be thought of as a phase that precedes our balancing process in which the players decide who they will partner with.

The algorithm is a max-product form of belief propagation. It was introduced for maximum matching on bipartite graphs by Bayati, Shah, and Sharma [4] and for maximum matching on general graphs by Bayati et al. [5] and independently by Sanghavi, Malioutov, and Willsky [19]. The algorithm is iterative; in each round every player sends a message to each of its neighbors. If $a_{u \to v}^t$ is the message that player $u$ sends player $v$ at time $t$, then $a_{u \to v}^t$ is computed recursively as follows:

$$a_{u \to v}^t = \max \left\{ \max_{q \sim u, q \neq v} \left\{ w_{q,u} - a_{q \to u}^{t-1} \right\}, 0 \right\} \ .$$

The messages are initialized to 0 at time 0. The algorithm ends when these messages converge to some fixed values $a_{u \to v}^*$ for all pairs of neighbors $(u,v)$ (if this does indeed happen). If the algorithm converges, then a matching $M$ is chosen where $\{u,v\} \in M$ if and only if $a_{u \to v}^* + a_{v \to u}^* \leq w_{uv}$. The main result of [5] and [19] is that this process converges if and only if the maximum fractional matching is integral (assuming that the edge weights have been perturbed so that the maximum fractional matching is unique).

In our network exchange context, we interpret the message $a_{u \to v}^t$ as the disagreement point that player $u$ has when bargaining with player $v$ over the value of an exchange on edge $uv$. Note that this message is computed in a way that is consistent with this interpretation: it is the maximum value $u$ can get by executing an exchange with a neighbor other than $v$ and offering that neighbor the same value that he is currently getting. (As before, $u$ does not have to make an exchange with any of its neighbors, so it can always guarantee itself a value of at least 0.) Also note that under this interpretation two players match with each other at convergence if and only if they have positive surplus. Finally, note that this process converges exactly in the case that makes sense in our context: when a stable outcome exists.

## B  Proof of Proposition 6

In this section, we prove Proposition 6 and show that the result is tight, to within a constant factor.

*Proof of Proposition 6.* Let $uv$ be the edge that maximizes $w_{u',v'} - x_{u'} - x_{v'}$ for all unmatched edges $u'v'$. Let $\delta = w_{u,v} - x_u - x_v$. To prove that $\mathbf{x}$ is $(n\varepsilon)$-stable, it suffices to show that $\delta \leq n\varepsilon$. Since $M$ is a maximum matching, at least one of $u$ and $v$ is matched. Without loss of generality, assume $u_0 = u$ is matched. Consider the exploration algorithm, and let $H \subseteq G$ be the vertices $u_\ell, \ldots, u_0, \ldots, u_r$ and edges traversed by the algorithm.

We first prove the following lemma.

**Lemma 9.** *Any edge $ab \in H \setminus M$ satisfies $x_a + x_b \leq w_{a,b} - \delta + (n-1)\varepsilon$.*

*Proof.* We prove a stronger claim: for $0 \leq i \leq \lfloor r/2 \rfloor$, $x_{2i-1} + x_{2i} \leq w_{2i-1,2i} - \delta + i\varepsilon$ (and a similar statement holds for $0 \geq i \geq \lceil (\ell+1)/2 \rceil$). The proof of this claim is by induction on $i$.

The base case is when $i = 0$. Since $u_0 \sim v$, we have $\alpha_0 = w_{-1,0} - x_{-1} \geq w_{0,v} - x_v = x_0 + \delta$. Hence, $x_{-1} + x_0 \leq w_{-1,0} - \delta$. (In fact, by the definition of $u = u_0$ and $v$, this implies that $x_{-1} + x_0 = w_{-1,0} - \delta$.)

Now suppose that the claim holds for $i-1$. That is, suppose that $x_{2i-3} + x_{2i-2} \leq w_{2i-3,2i-2} - \delta + (i-1)\varepsilon$. Since $u_{2i-3} \sim u_{2i-2}$, it follows that $\alpha_{2i-2} \geq w_{2i-3,2i-2} - x_{2i-3}$. Thus, we have

$$\begin{aligned}
x_{2i-1} + x_{2i} &= x_{2i-1} + w_{2i-1,2i} - \alpha_{2i-1} & \text{(by choice of } u_{2i}\text{)} \\
&\leq w_{2i-1,2i} + x_{2i-2} - \alpha_{2i-2} + \varepsilon & (\varepsilon\text{-quasi-balanced}) \\
&\leq w_{2i-1,2i} + x_{2i-2} - w_{2i-3,2i-2} + x_{2i-3} + \varepsilon & \text{(from above)} \\
&\leq w_{2i-1,2i} - \delta + i\varepsilon & \text{(IH),}
\end{aligned}$$

as desired.

By an analogous argument, we can show that for $0 \geq i \geq \lceil(\ell+1)/2\rceil$, we have $x_{2i-1} + x_{2i} \leq w_{2i-1,2i} - \delta + i\varepsilon$. Hence, every edge $ab \in H \setminus M$ satisfies $x_a + x_b \leq w_{ab} - \delta + \max\{|\lceil(\ell+1)/2\rceil|, \lfloor r/2 \rfloor\}\varepsilon \leq w_{ab} - \delta + (n-1)\varepsilon$. $\square$

With the lemma proved, we consider the possible types of structures $H$ (see Figure 2). Since $M$ is a maximum matching, we can use the standard matching results shown in Appendix C to show that $\delta \leq n\varepsilon$ in every case as follows:

- Suppose that $H$ is capped (Figure 2(a)). Without loss of generality, assume $u_{r-1}u_r$ is matched. Hence, $\alpha_r = 0$, and $x_r - \alpha_r = x_r \geq 0$. Because $u_{r-1}u_r$ is $\varepsilon$-balanced, $x_{r-1} - \alpha_{r-1} \geq -\varepsilon$. Since $u_{r-1} \sim u_{r-2}$, we have $\alpha_{r-1} \geq w_{r-2,r-1} - x_{r-2}$. From this and Lemma 9, we have
$$0 \leq x_{r-1} - \alpha_{r-1} + \varepsilon \leq x_{r-1} - w_{r-2,r-1} + x_{r-2} + \varepsilon \leq -\delta + n\varepsilon \ ,$$
and hence $\delta \leq n\varepsilon$ as desired.

- Suppose that $H$ is a lollypop (Figure 2(b)) or a pretzel (Figure 2(f)). If it is the latter, then a simple parity argument shows that one of the cycles of $H$ must have even length. Thus in either case, $H$ contains an even alternating cycle. Relabel the vertices of this cycle $v_0, v_1, \ldots, v_{2k}$, where $v_{2k} = v_0$ and $v_0v_1 \in M$. By Lemma 9 and Lemma 10, we have
$$\sum_{i=1}^{k}(x_{2i-1} + x_{2i}) \leq \sum_{i=1}^{k} w_{2i-1,2i} - k\delta + kn\varepsilon \leq \sum_{i=0}^{k-1} w_{2i,2i+1} - k\delta + kn\varepsilon \ .$$

  On the other hand,
  $$\sum_{i=1}^{k}(x_{2i-1} + x_{2i}) = \sum_{i=0}^{k-1}(x_{2i} + x_{2i+1}) = \sum_{i=0}^{k-1} w_{2i,2i+1} \ .$$

  Combining the above and dividing by $k$ yields $\delta \leq n\varepsilon$ as desired.

- Suppose that $H$ is an augmenting path (Figure 2(c)). Relabel the vertices of the augmenting path $v_0, v_1, \ldots, v_{2k+1}$. By Lemma 9 and Lemma 10, we have
$$\begin{aligned}
\sum_{i=0}^{k}(x_{2i} + x_{2i+1}) &\leq \sum_{i=0}^{k} w_{2i,2i+1} - (k+1)\delta + (k+1)n\varepsilon \\
&\leq \sum_{i=1}^{k} w_{2i-1,2i} - (k+1)\delta + (k+1)n\varepsilon \ .
\end{aligned}$$

  On the other hand, since $x_0 = x_{2k+1} = 0$, we have
  $$\sum_{i=0}^{k}(x_{2i} + x_{2i+1}) = \sum_{i=1}^{k}(x_{2i-1} + x_{2i}) = \sum_{i=1}^{k} w_{2i-1,2i} \ .$$

  Combining the above and dividing by $k+1$ yields $\delta \leq n\varepsilon$ as desired.

- Suppose that $H$ is a flower (Figure 2(d)). Relabel the vertices of the flower starting at the bottom of the stem by $v_0, v_1, \ldots, v_{2k}, \ldots, v_{2\ell+1}$, where $v_{2k} = v_{2\ell+1}$. By Lemma 9 and Lemma 11, we know

$$
\begin{aligned}
& 2\sum_{i=0}^{k-1}(x_{2i} + x_{2i+1}) + \sum_{i=k}^{\ell}(x_{2i} + x_{2i+1}) \\
\leq\ & 2\sum_{i=0}^{k-1} w_{2i,2i+1} + \sum_{i=k}^{\ell} w_{2i,2i+1} - (k+\ell+1)\delta + (k+\ell+1)n\varepsilon \\
\leq\ & 2\sum_{i=1}^{k} w_{2i-1,2i} + \sum_{i=k+1}^{\ell} w_{2i-1,2i} - (k+\ell+1)\delta + (k+\ell+1)n\varepsilon \ .
\end{aligned}
$$

On the other hand, since $x_0 = 0$ and $x_{2k} = x_{2\ell+1}$, we have

$$
\begin{aligned}
2\sum_{i=0}^{k-1}(x_{2i} + x_{2i+1}) + \sum_{i=k}^{\ell}(x_{2i} + x_{2i+1}) &= 2\sum_{i=1}^{k}(x_{2i-1} + x_{2i}) + \sum_{i=k+1}^{\ell}(x_{2i-1} + x_{2i}) \\
&= 2\sum_{i=1}^{k} w_{2i-1,2i} + \sum_{i=k+1}^{\ell} w_{2i-1,2i} \ .
\end{aligned}
$$

Combining the above and dividing by $(k+\ell+1)$ yields $\delta \leq n\varepsilon$ as desired.

- Finally, suppose that $H$ is a bicycle. Relabel the vertices by

$$v_0, \ldots, v_{2j+1}, \ldots, v_{2k}, \ldots, v_{2\ell+1} \ ,$$

such that $v_0 = v_{2j+1}$ and $v_{2k} = v_{2\ell+1}$. By Lemma 9 and Lemma 12, we have

$$
\begin{aligned}
& 2\sum_{i=j+1}^{k-1}(x_{2i} + x_{2i+1}) + \sum_{i=0}^{j}(x_{2i} + x_{2i+1}) + \sum_{i=k}^{\ell}(x_{2i} + x_{2i+1}) \\
\leq\ & 2\sum_{i=j+1}^{k-1} w_{2i,2i+1} + \sum_{i=0}^{j} w_{2i,2i+1} + \sum_{i=k}^{\ell} w_{2i,2i+1} - (\ell+k-j)\delta + (\ell+k-j)n\varepsilon \\
\leq\ & 2\sum_{i=j+1}^{k} w_{2i-1,2i} + \sum_{i=1}^{j} w_{2i-1,2i} + \sum_{i=k+1}^{\ell} w_{2i-1,2i} - (\ell+k-j)\delta + (\ell+k-j)n\varepsilon \ .
\end{aligned}
$$

On the other hand,

$$
\begin{aligned}
& 2\sum_{i=j+1}^{k-1}(x_{2i} + x_{2i+1}) + \sum_{i=0}^{j}(x_{2i} + x_{2i+1}) + \sum_{i=k}^{\ell}(x_{2i} + x_{2i+1}) \\
=\ & 2\sum_{i=j+1}^{k}(x_{2i-1} + x_{2i}) + \sum_{i=1}^{j}(x_{2i-1} + x_{2i}) + \sum_{i=k+1}^{\ell}(x_{2i-1} + x_{2i}) \\
=\ & 2\sum_{i=j+1}^{k} w_{2i-1,2i} + \sum_{i=1}^{j} w_{2i-1,2i} + \sum_{i=k+1}^{\ell} w_{2i-1,2i} \ .
\end{aligned}
$$

Combining the above and dividing by $(\ell+k-j)$ yields $\delta \leq n\varepsilon$ as desired.

Thus, in all cases, we have shown that $\delta \leq n\varepsilon$, and hence $\mathbf{x}$ is $(n\varepsilon)$-stable. $\square$

Note that this result is tight to within a constant factor even for unweighted graphs. We present a family of graphs proving this fact: Consider an unweighted lollypop consisting of $2k+1$ vertices $v_0, u_1, v_1, u_2, v_2, \ldots, u_k, v_k$ where $u_1, v_1, \ldots, u_k, v_k$ form an alternating cycle, and $v_0$ is adjacent only to $u_1$. (Hence $u_j v_j$ is matched for $1 \leq j \leq k$.) Let $\varepsilon = 4/(k^2 + 4k + 4)$, and consider the allocation $\mathbf{x}$ given by $x_{u_j} = 1 - j(k+2-j)\varepsilon$ and $x_{v_j} = j(k+2-j)\varepsilon$ for $1 \leq j \leq k$. It can be checked that every vertex gets an amount between 0 and 1, so this is a valid allocation. It can also be checked that this allocation is $(2\varepsilon)$-quasi-balanced. However, $x_{v_0} + x_{u_1} = x_{u_1} = 1 - (k+1)\varepsilon$, and therefore this allocation is not $\delta$-stable for any $\delta = o(n\varepsilon)$.

## C  Structural Lemmas on Maximum Matchings

For completeness, we present proofs for some known structural results on maximum matchings [2, 14]. The pertinent structures are shown in Figure 2 and Figure 3.

**Lemma 10.** *Let $G$ be a graph and $M$ a maximum weighted matching on $G$. Let $H \subseteq G$ be an alternating cycle, or an alternating path whose endpoints are not matched outside of $H$. Then $w(H \setminus M) \leq w(H \cap M)$.*

*Proof.* Let $M' = M \triangle H$. Since $H$ is either a cycle, or its endpoints are not matched outside of $H$, we know $M'$ is a valid matching. The lemma follows because $w(M) \geq w(M')$. □

**Lemma 11.** *Let $M$ be a maximum matching on $G$; suppose that $M$ is also a maximum fractional matching; and let $H = u_0, u_1, \ldots, u_{2k}, \ldots, u_{2\ell+1}$ be the vertices of a flower (in order), where $u_{2k} = u_{2\ell+1}$. Then*

$$2 \sum_{i=0}^{k-1} w_{2i,2i+1} + \sum_{i=k}^{\ell} w_{2i,2i+1} \leq 2 \sum_{i=1}^{k} w_{2i-1,2i} + \sum_{i=k+1}^{\ell} w_{2i-1,2i} \ .$$

*Proof.* Consider the fractional matching $\mathbf{y}$ given by

$$y_e = \begin{cases} 1 & \text{if } e \in M \setminus (M \cap H) \\ 1 & \text{if } e = u_{2i} u_{2i+1} \text{ for } 0 \leq i \leq k-1 \\ \frac{1}{2} & \text{if } e = u_i u_{i+1} \text{ for } 2k \leq i \leq 2\ell \\ 0 & \text{otherwise} \end{cases} .$$

Because $M$ is a maximum fractional matching, $2w(M) \geq 2w(\mathbf{y})$. Hence,

$$2 \sum_{i=1}^{k} w_{2i-1,2i} + 2 \sum_{i=k+1}^{\ell} w_{2i-1,2i} \geq 2 \sum_{i=0}^{k-1} w_{2i,2i+1} + \sum_{i=2k}^{2\ell} w_{i,i+1} \ ,$$

which can be rearranged to yield the lemma. □

**Lemma 12.** *Let $M$ be a maximum matching on $G$; suppose that $M$ is also a maximum fractional matching; and let $H = u_0, \ldots, u_{2j+1}, \ldots, u_{2k}, \ldots, u_{2\ell+1}$, be the vertices of a bicycle (in order), where $u_0 = u_{2j+1}$ and $u_{2k} = u_{2\ell+1}$. Then*

$$2 \sum_{i=j+1}^{k-1} w_{2i,2i+1} + \sum_{i=0}^{j} w_{2i,2i+1} + \sum_{i=k}^{\ell} w_{2i,2i+1}$$
$$\leq \ 2 \sum_{i=j+1}^{k} w_{2i-1,2i} + \sum_{i=1}^{j} w_{2i-1,2i} + \sum_{i=k+1}^{\ell} w_{2i-1,2i} \ .$$

*Proof.* Consider the fractional matching $\mathbf{y}$ given by

$$y_e = \begin{cases} 1 & \text{if } e \in M \setminus (M \cap H) \\ 1 & \text{if } e = u_{2i} u_{2i+1} \text{ for } j+1 \leq i \leq k-1 \\ \frac{1}{2} & \text{if } e = u_i u_{i+1} \text{ for } 0 \leq i \leq 2j \\ \frac{1}{2} & \text{if } e = u_i u_{i+1} \text{ for } 2k \leq i \leq 2\ell \\ 0 & \text{otherwise} \end{cases} .$$

Because $M$ is a maximum fractional matching, $2w(M) \geq 2w(\mathbf{y})$. Hence,

$$2 \sum_{i=j+1}^{k} w_{2i-1,2i} + 2 \sum_{i=1}^{j} w_{2i-1,2i} + 2 \sum_{i=k+1}^{\ell} w_{2i-1,2i}$$
$$\geq 2 \sum_{i=j+1}^{k-1} w_{2i,2i+1} + \sum_{i=0}^{2j} w_{i,i+1} + \sum_{i=k}^{2\ell} w_{i,i+1} ,$$

which can be rearranged to yield the lemma. □

Lemmas 10, 11, and 12 allow us to prove Lemma 7, from Section 3.

*Proof of Lemma 7.* Suppose that $H$ is an augmenting path. Label the vertices in this path $u_0, u_1, \ldots, u_{2k+1}$. Since $u_0$ and $u_{2k+1}$ are unmatched, we have $x_0 = x_{2k+1} = 0$, and

$$\sum_{i=0}^{k} w_{2i,2i+1} > \sum_{i=0}^{k}(x_{2i} + x_{2i+1}) = \sum_{i=1}^{k}(x_{2i-1} + x_{2i}) = \sum_{i=1}^{k} w_{2i-1,2i} .$$

Hence by Lemma 10, $M$ is not maximum.

Now suppose that $H$ is an alternating cycle. Label the vertices in this cycle $u_0, u_1, \ldots, u_{2k-1}$ such that $u_0 u_1 \in M$. Also, define $u_{2k}$ to be the vertex $u_0$. We have

$$\sum_{i=1}^{k} w_{2i-1,2i} > \sum_{i=1}^{k}(x_{2i-1} + x_{2i}) = \sum_{i=0}^{k-1}(x_{2i} + x_{2i+1}) = \sum_{i=0}^{k-1} w_{2i,2i+1} .$$

Hence, by Lemma 10, $M$ is not maximum.

Now suppose that $H$ is a flower. Label the vertices in the stem of the flower by $u_0, u_1, \ldots, u_{2k}$ and label the vertices in the blossom by $u_{2k}, u_{2k+1}, \ldots, u_{2\ell+1}$ (where $u_{2\ell+1} = u_{2k}$). Since $x_0 = 0$, we have

$$2\sum_{i=0}^{k-1} w_{2i,2i+1} + \sum_{i=k}^{\ell} w_{2i,2i+1} > 2\sum_{i=0}^{k-1}(x_{2i} + x_{2i+1}) + \sum_{i=k}^{\ell}(x_{2i} + x_{2i+1})$$
$$= 2\sum_{i=1}^{k}(x_{2i-1} + x_{2i}) + \sum_{i=k+1}^{\ell}(x_{2i-1} + x_{2i})$$
$$= 2\sum_{i=1}^{k} w_{2i-1,2i} + \sum_{i=k+1}^{\ell} w_{2i-1,2i} .$$

Hence, by Lemma 11, $M$ is not a maximum fractional matching.

Finally, suppose that $H$ is a bicycle. Label the vertices in the first blossom by $u_0, u_1, \ldots, u_{2j+1}$ (where $u_{2j+1} = u_0$); label the vertices in the stem by $u_{2j+1}, u_{2j+2}, \ldots, u_{2k}$; and label the vertices in the second blossom by $u_{2k}, u_{2k+1}, \ldots, u_{2\ell+1}$ (where $u_{2\ell+1} = u_{2k}$). We have

$$2\sum_{i=j+1}^{k-1} w_{2i,2i+1} + \sum_{i=0}^{j} w_{2i,2i+1} + \sum_{i=k}^{\ell} w_{2i,2i+1}$$
$$> 2\sum_{i=j+1}^{k-1}(x_{2i} + x_{2i+1}) + \sum_{i=0}^{j}(x_{2i} + x_{2i+1}) + \sum_{i=k}^{\ell}(x_{2i} + x_{2i+1})$$
$$= 2\sum_{i=j+1}^{k}(x_{2i-1} + x_{2i}) + \sum_{i=1}^{j}(x_{2i-1} + x_{2i}) + \sum_{i=k+1}^{\ell}(x_{2i-1} + x_{2i})$$
$$= 2\sum_{i=j+1}^{k} w_{2i-1,2i} + \sum_{i=1}^{j} w_{2i-1,2i} + \sum_{i=k+1}^{\ell} w_{2i-1,2i} .$$

Hence, by Lemma 12, $M$ is not a maximum fractional matching. □